\documentclass[%
 reprint,
 amsmath,amssymb,
 aps,
]{revtex4-2}
\usepackage{graphicx}
\usepackage{dcolumn}
\usepackage{bm}
\usepackage[usenames]{color}
\usepackage{mathrsfs}
\usepackage{colortbl}
\usepackage{hyperref}
\begin{document}

\title{Intrinsic Nonreciprocity in Asymmetric Josephson Junctions with Non-Sinusoidal Current-Phase Relations}
\author{R.A. Hovhannisyan}
\email{razmik.hovhannisyan@fysik.su.se}
\affiliation{Department
of Physics, Stockholm University, AlbaNova University Center,
SE-10691 Stockholm, Sweden} 

\begin{abstract}
Josephson junctions (JJs) with non-sinusoidal current-phase relations (CPRs) have gathered increasing attention, partly due to growing interest in topological 2D materials. Understanding how CPR and inhomogeneities in JJs influence their response is crucial for accurate interpretation of experimental observations.
This Letter reports that a non-sinusoidal CPR, combined with asymmetries in the JJ, can break spatial symmetry and give rise to the Josephson diode effect (JDE) in the short junction regime. This nonreciprocity is shown to emerge as an intrinsic mechanism related to the maximization of the supercurrent, rather than being solely driven by geometric or material asymmetries. Further analysis shows that JDE efficiency is strongly influenced by the CPR shape but is largely insensitive to junction asymmetry, making the observed nonreciprocity not only a potential experimental signature of unconventional CPRs but also a possible method for probing their properties.

\end{abstract}

\maketitle



JJs are characterized by their highly nonlinear response to external stimuli, enabling a range of unique phenomena in superconducting electronics~\cite{Barone,Likharev_1979,RSFQ,Bairamkulov_2025}. One of the key characteristics of these devices is the CPR. While the CPR was originally derived to be sinusoidal~\cite{Larkin_1969,Josepshon_1964}, it is now well established that real junctions can exhibit significant deviations from these simpler forms~\cite{Kulik_Clean_1,Kulik_Clean_2,Kulik_Dirt,Kulik_Phase,Golubov_Review,Zaikin,Ivanov_1981,Kupr_1992,Buzdin_1982,Bezryadin_2008,Troeman_2008,Nanda_2017,Stolyarov_2020,Karter_2015,Spanton_2017,Hart_2019,Frolov_2004,Sochnikov_2013,Kayyalha_2020,Rocca_2007,Ginzburg_2018,Enders_2023,Babich_2023}.

The properties of JJs with non-sinusoidal CPRs under homogeneous magnetic fields and uniform critical current density distributions $J_c(x)$ have been well investigated to date, both experimentally~\cite{Stolyarov_2020,Frolov_2004,Willsch_2024} and theoretically~\cite{Golubov_Review,Goldobin_2007}.  However, in many experimental situations, these ideal conditions are not reproduced due to nonhomogeneities in geometry of the junctions~\cite{Barone,Raz_2024}, quantum Hall effect~\cite{Hall_1,Hall_2}, edge states~\cite{Chen_2024}, etc. Although such situations have been considered previously, to my knowledge, no substantial analysis has been conducted on the impact of inhomogeneous current density distributions on the properties of JJs, especially in the long-junction limit.

Another motivation for this study stems from the growing interest in superconducting junctions based on 2D materials~\cite{Bezryadin_2008,Troeman_2008,Nanda_2017,Stolyarov_2020,Karter_2015,Spanton_2017,Hart_2019,Frolov_2004,Sochnikov_2013,Kayyalha_2020,Rocca_2007,Ginzburg_2018,Enders_2023,Babich_2023},  which often exhibit unconventional CPRs. Recently, Ref.~\cite{Fominov_2022} demonstrated the appearance of the diode effect in an asymmetric SQUID with a non-sinusoidal current-phase relation (CPR). Moreover, Ref.~\cite{Kudriashov_2025} reported the experimental observation of non-reciprocal behavior resulting from an asymmetric critical current ($J_c$) in JJs. Nevertheless, the JDE in 2D materials can also arise from a variety of other mechanisms, including spin-orbit interactions (SOI)~\cite{Wakatsuki1,Wakatsuki2} or the intrinsic lack of inversion symmetry in the crystal structure~\cite{Kim_2024}. Additionally, external factors like applied magnetic fields~\cite{Hoshino_2018} or proximity effects in hybrid heterostructures~\cite{Prox} may also contribute to the emergence of nonreciprocal superconducting transport (for more, see the review~\cite{Reveiew_diode} and references therein). Therefore, the ability to distinguish an asymmetric $J_c$ from other mechanisms is crucial to correctly interpret the observed nonreciprocal behavior.

In this Letter, I systematically analyze the influence of non-sinusoidal CPRs on the magnetic field dependence of the critical current \( I_c(H) \) of JJ. By exploring both short and long junction regimes, I find that asymmetries in the critical current density \( J_c(x) \), when combined with non-sinusoidal CPRs, lead to nonreciprocal behavior in the short-junction limit. Through a straightforward theoretical analysis, I show that in this regime, nonreciprocity arises as an intrinsic consequence of the critical current maximization mechanism. Further analysis of the JDE, considering both the asymmetry of the junction and the form of the CPR, reveals that while the maximum diode efficiency \( \eta \) is highly sensitive to the shape of the CPR, it can appear in JJs with relatively small asymmetry and is only weakly affected by its degree. 
This contrast in sensitivity may offer a practical approach for extracting CPR characteristics from measurements of JDE in asymmetric JJs.



\begin{figure*}[ht!]
    \begin{center}
    \includegraphics[width =1.99\columnwidth]{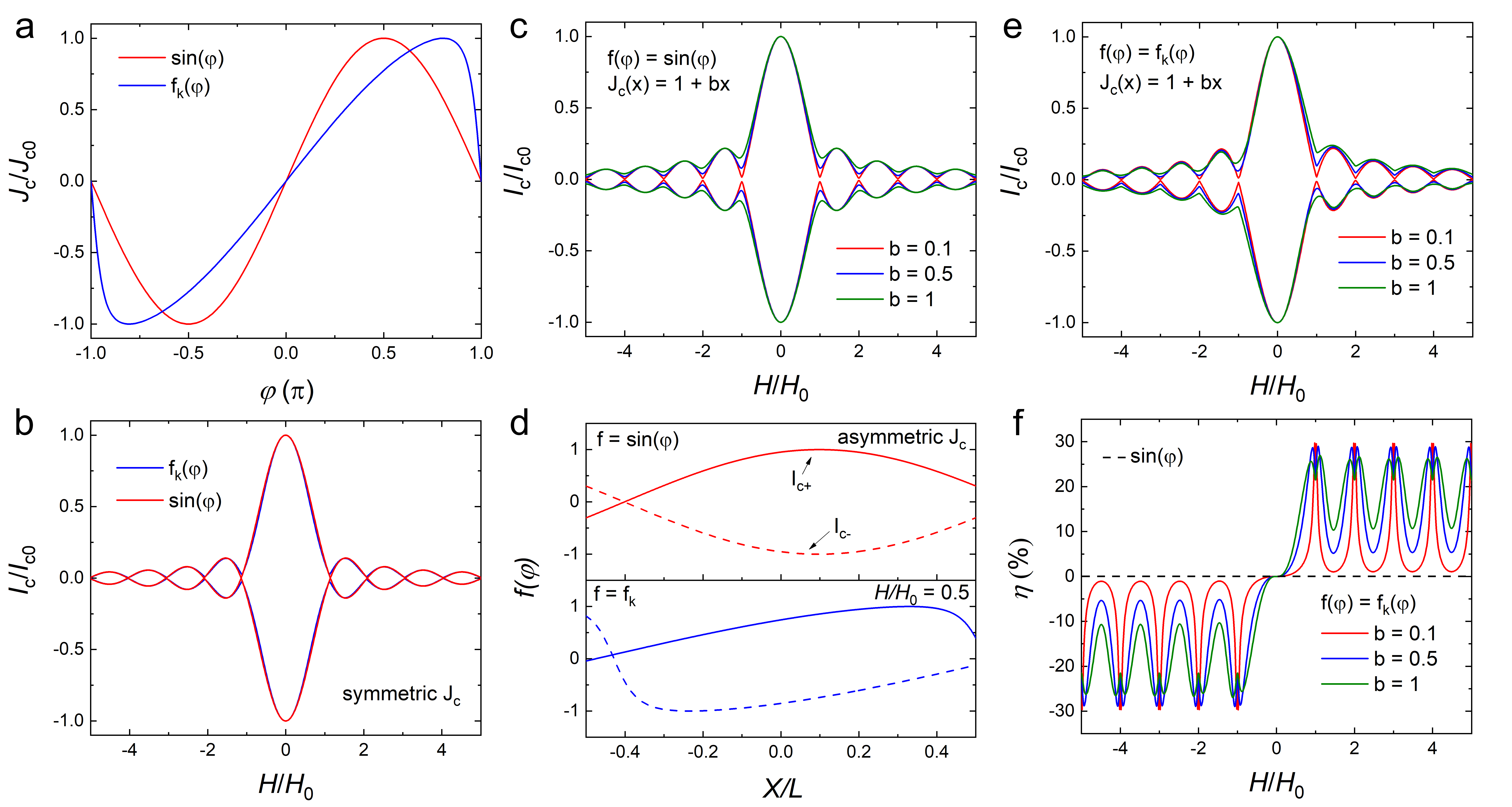}
    \caption{\textbf{Diode effect in a JJ with a nonuniform distribution of current density and a nontrivial current-phase relation.} 
(a) CPRs used in the calculations: the sinusoidal CPR described by Eq.~\ref{sinphi} (red curve) and the non-sinusoidal CPR given by Eq.~\ref{kulik} (blue curve)  corresponding to a transparency parameter \( D = 0.99 \). 
(b) Normalized critical current \( I_c \) of the JJ as a function of applied magnetic field, computed for a symmetric critical current density profile \( J_c(x) = 1 + x^2/L^2 \) using the CPRs shown in (a). 
(c) \( I_c(H) \) dependence for an asymmetric critical current density \( J_c(x) = 1 + bx/L \) and a sinusoidal CPR. The curves demonstrate the bigger node lifting with increasing \( b \), while preserving both time-reversal and spatial symmetry. 
(d) spatial distribution of \( f(\varphi) \) for \( I_{c+} \) (solid line) and \( I_{c-} \) (dashed line) at a magnetic field corresponding to half a flux quantum through the junction, i.e., \( H/H_0 = 0.5 \), and \( b = 0.5 \).
(e) Critical current \( I_c \) as a function of magnetic field for the non-sinusoidal CPR as in (a), and an asymmetric \( J_c(x) \) as in (c). The JJ exhibits broken spatial symmetry while preserving time-reversal symmetry.
(f) Diode efficiency corresponding to the \( I_c(H) \) curves in (e), evaluated as a function of the applied magnetic field. Dashed black curve demonstrate absence of JDE in the case of (c).
}
\label{Fig:1}
    \end{center}
\end{figure*}

I begin by considering the short-junction limit, where the size of the Josephson junction can be considered to be smaller than the Josephson penetration depth, i.e. \( L \ll \lambda_J \). In this case, one can neglect the screening of the magnetic field by JJ, and the phase shift originating from it can be expressed by the relation:

\begin{equation}\label{shortPhase}
    \varphi = \frac{2\pi d_{\mathrm{eff}}}{\Phi_0} H + \phi
\end{equation}
where \( \Phi_0 = H_0d_{eff}L \) is the magnetic flux quantum, \( d_{\mathrm{eff}} \) is the effective magnetic thickness of the junction, and \( H \) is the external magnetic field. The term \( \phi \) in Eq.~\ref{shortPhase} is an integration constant.

The critical current is then expressed as:

\begin{equation}\label{SingleJJ}
  I_s(H)
  = \max_{\phi}\left[
    \int\limits_{-L/2}^{L/2} 
    J_c(x)\, 
    f(\varphi,x)
    \,\mathrm{d}x
  \right],
\end{equation}
where $J_c$ is the critical current density distribution and  \( f(\varphi,x) \) denotes the appropriate CPR, which can take an arbitrary form depending on the properties of the junction~\cite{Golubov_Review}. However, for the sake of simplicity, the spatial distribution of the CPR will be neglected.

For further analysis, we focused on two main types of CPRs Fig.~\ref{Fig:1}(a) and their respective analyses. The first one is the ordinary sinusoidal relation:
\begin{equation}~\label{sinphi}
    f = \sin \varphi.
\end{equation}

The second one corresponds to the clean limit of an narrow SNS-type junction~\cite{Kulik_Phase}:
\begin{equation}\label{kulik}
    f_k = \frac{\sin 
    \varphi}{\sqrt{1 - D\sin^2(\varphi/2)}},  
\end{equation}  
where \( D \) is the transparency parameter.

Fig.~\ref{Fig:1}(b) demonstrates the calculated $I_c(H)$ dependencies for a symmetric distribution of $J_c(x) = 1 + x^2/L^2$ (edge state-induced inhomogeneity) for both sinusoidal (Eq.~\ref{sinphi}) and non-sinusoidal (Eq.~\ref{kulik}) CPRs with transparency parameter D = 0.99. It can be observed that, in both cases, the impact of the form of CPR has a negligible effect on the overall shape of the critical current dependence. During the simulation, we tested a variety of symmetric $J_c(x)$ and $f(\varphi)$ dependencies, and no significant difference was observed in correspondence with previous experimental results. Moreover, the critical current in the negative and positive direction are equal in amplitude $I_{c+} = I_{c-}$ showcasing no evidence of JDE. 

In contrast, Fig.~\ref{Fig:1}(c) and Fig.~\ref{Fig:1}(e) present numerically calculated \( I_c(H) \) dependencies for an asymmetrical linear distribution of the critical current density, given by \( J_c(x) = 1 + bx/L \) (gradient-induced inhomogeneity) , for different values of the tilt parameter \( b \). 

First, it can be observed that in both cases, the critical current dependencies exhibit node lifting, in contrast to the symmetrical \( J_c(x) \) Fig.~\ref{Fig:1}(b). Furthermore, for \( f(x) = \sin x \) (Fig.~\ref{Fig:1}(c)), the critical currents \( I_{c+} \) and \( I_{c-} \) remain equal for any applied magnetic field, indicating the preservation of both spatial and time-reversal symmetries. Indeed, if any \( \phi = \varphi_0 \) maximizes Eq.~\ref{SingleJJ}, then \( \varphi_0 + \pi \) minimizes it (see Fig.~\ref{Fig:1}(d), top panel, for the spatial distribution of the phase at an applied magnetic field equal to half a flux quantum).

On the other hand, Fig.~\ref{Fig:1}(e) reveals a notable asymmetry in both the critical currents $I_{c+}\neq I_{c-}$ and spatial distribution of phase within the length of the junction, as shown on Fig.~\ref{Fig:1}(d) bottom panel. 
This leads to a  JDE with efficiency:
\begin{equation}
   \eta = \frac{I_{c+} - I_{c-}}{I_{c+} + I_{c-}},  
\end{equation}  
as demonstrated in Fig.~\ref{Fig:1}(f).

\begin{figure*}[ht!]
    \begin{center}
    \includegraphics[width =1.99\columnwidth]{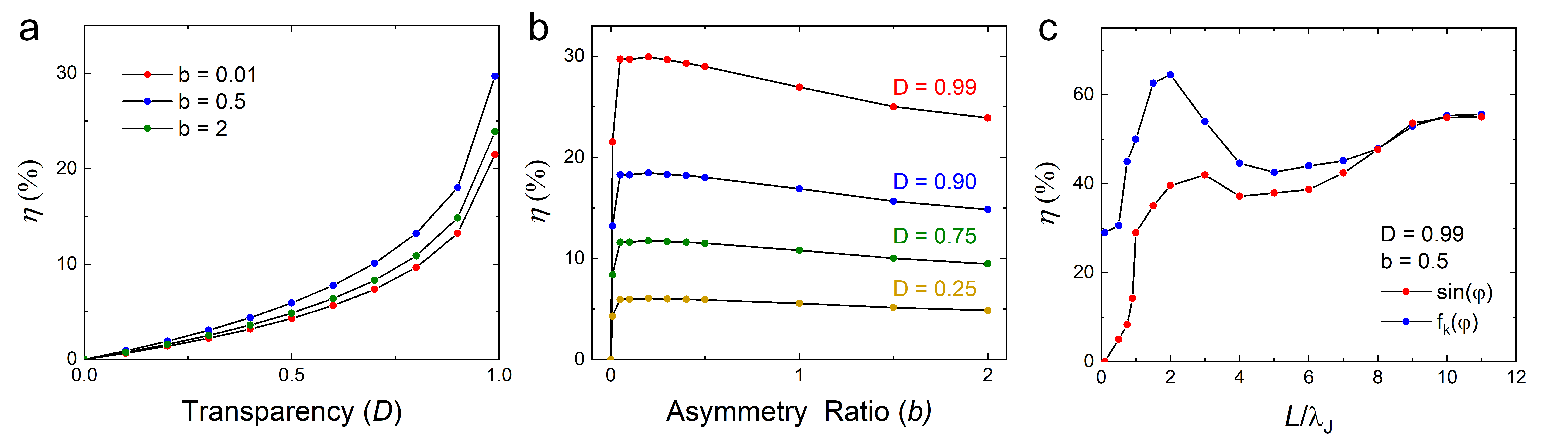}\caption{
  \textbf{Dependence of maximum JDE on different junction parameters} 
  (a) Diode efficiency $\eta$ as a function of the CPR transparency parameter \( D \) (Eq.~\ref{kulik}). A strong increase in $\eta$ is observed with higher transparency, highlighting the significant impact of the CPR form on nonreciprocity.  
  (b) $\eta$ as a function of the asymmetry parameter \( b \) in the critical current density distribution \( J_c(x) = 1 + bx/L \). While a step-like increase in $\eta$ is visible for small values of \( b \), the overall sensitivity to asymmetry is weak.  
  (c) Diode efficiency $\eta$ versus normalized junction length \( L/\lambda_J \). In the short-junction regime (\( L/\lambda_J \ll 1 \)), the non-sinusoidal CPR exhibits residual JDE due to intrinsic mechanisms, while in the long-junction regime (\( L/\lambda_J > 9 \)), both CPRs demonstrate asymptotic diode efficiency ($\sim$ 55\%) driven by self-field (ratchet) effects. The crossover region (\( 1 < L/\lambda_J < 4 \)) reveals the complex interplay between intrinsic and self-field mechanisms.
}\label{Fig:2}
    \end{center}
\end{figure*}

To further investigate the appearance of nonreciprocity in asymmetrical junctions, we have calculated the maximum efficiency, $\eta$, reached within the vicinity of the first minimum of the $I_c(H)$ pattern as a function of the CPR (transparency parameter $D$ in Eq.~\ref{kulik}) and the asymmetry ratio $b$. Figure~\ref{Fig:2} (a) and (b) present the results of these calculations respectively.  It can be observed that the efficiency exhibits a strong exponential-like dependence on the CPR. Interestingly, unlike the CPR, the maximum achieved nonreciprocity shows a low sensitivity to the asymmetry of $J_c(x)$, displaying a step-like rise in $\eta$ within the range of $b < 0.05$ and a slowly decaying tail as the total asymmetry increases. Note that the asymmetry of \( J_c(x) \) strongly affects the overall behavior of \( \eta(H) \) with respect to the magnetic field (see Fig.~\ref{Fig:1}(f)).

Finally, we calculated the diode efficiency dependence in the finite-length junction limit using the sine-Gordon formalism~\cite{Krasnov_In} for both CPRs. The results are shown in Fig.~\ref{Fig:2}(c). In contrast to the short junction limit, both CPRs now exhibit significant $\eta$, arising from the self-field effects of the JJ, this phenomenon is also known as the vortex ratchet effect~\cite{Krasnov_Diode,Vilegas_2003,Moschalkov_2006,Lustikova_2018}. For sufficiently long junctions ($L/\lambda_J > 9$), both CPRs converge to an asymptotic efficiency of $\eta = 55\%$. However, they start to diverge at smaller $L/\lambda_J$, becoming evident for $L/\lambda_J < 4$. The $\sin(\varphi)$ curve peaks at $\eta \simeq 40\%$, while $f_k$ reaches a significantly higher $\eta \simeq 65\%$.  As expected, the efficiency of the sinusoidal CPR vanishes as $L \to 0$, while the other retains a residual nonreciprocity of $\eta \simeq 30\%$, consistent with Fig.~\ref{Fig:2}(a) and (b).


Thus, we investigate properties of JJs with a nonuniform critical current density \( J_c(x) \) and different types of CPRs. To understand the cause of the observed nonreciprocity in the short-junction regime, a careful analysis of Eqs.~\ref{shortPhase}-~\ref{kulik} is needed. In general, the CPR of a JJ can be written as a sum of Fourier components.
\begin{equation}\label{eqSum}
    f = \sum_{n=1}^{\infty} a_n \sin{(n \varphi)}
\end{equation}
where \( a_n \) represents the amplitude of each individual Fourier harmonic. 

Eq.~\ref{SingleJJ} can be solved analytically under the assumption that the discrete Fourier transform of \( J_c(x) \) are known for the frequencies defined in terms of the external magnetic field \( H \):
\begin{equation}~\label{An}
    A_n = \int\limits_{-L/2}^{L/2} J_c(x) \cos{\left(\frac{2\pi d_{\mathrm{eff}}}{\Phi_0} nHx\right)} \, \mathrm{d}x,
\end{equation}
\begin{equation}~\label{Bn}
    B_n = \int\limits_{-L/2}^{L/2} J_c(x) \sin{\left(\frac{2\pi d_{\mathrm{eff}}}{\Phi_0} nHx\right)} \, \mathrm{d}x.
\end{equation}
Substituting Eq.~\ref{eqSum} and the expressions for \( A_n \) and \( B_n \) into Eq.~\ref{SingleJJ}, one obtains:
\begin{equation}\label{Eq_IcSum}
    I_c(H) = \sum_{n=1}^{\infty} \left( a_nA_n \sin{(n\phi)} + a_nB_n \cos{(n\phi)} \right).
\end{equation}
Furthermore, maximization with respect to the Josephson free phase \( \phi \) reduces to solving a trigonometric equation:
\begin{equation}\label{final}
     \sum_{n=1}^{\infty} \left( na_nA_n \cos{(n\phi)} - na_nB_n \sin{(n\phi)} \right) = 0.
\end{equation}
In typical experiments, higher-order harmonics can be neglected. In this case, Eq.~\ref{final} is simplified to solving a \( n \)-th order polynomial equation~\cite{Goldobin_2007}.


Now the appearance of Josephson diode effect becomes evident. In the case of a symmetric \( J_c \) (i.e., an even function of \( x \)), the integrand in Eq.~\ref{Bn} becomes an odd function, and therefore \( B_n = 0 \). In addition, \( I_c(\phi) \), given by Eq.~\ref{Eq_IcSum}, is an odd function of \( \phi \), satisfying \( I_c(\phi) = -I_c(-\phi) \). Thus, if \( \phi = \varphi_0 \) maximizes the critical current, then \( \phi = -\varphi_0 \) minimizes it. Furthermore, the integral in Eq.~\ref{An}  vanishes for any \( H = k\Phi_0 / L d_{\text{eff}} \) due to the inherent \( 2\pi \)-periodicity of the integrand. As a result, no node lifting of critical current can occur under these conditions.

This reasoning becomes invalid in the case of an asymmetric \( J_c(x) \), as the assumption of even symmetry no longer holds. This may lead to nonzero values of \( B_n \), and consequently, to an asymmetric \( I_c(\phi) \). Such asymmetry in \( I_c(\phi) \) can result in both node lifting and the emergence of JDE. Note that these arguments are not valid for a sinusoidal CPR or any other CPR with translational symmetry, i.e., \( f(\varphi) = - f(\varphi + a) \). 

Importantly, this mechanism of nonreciprocity is fundamentally different from that in the long-junction regime, as it does not involve self-field effects. This phenomenon is evident from the fact that, unlike in long junctions, the main maximum of \( I_c(H) \) in the short-junction regime appears at \( H = 0 \) Fig.~\ref{Fig:1}(c,e). 
The simultaneous presence of nonreciprocity and the absence of self-field effects can be interpreted as a unique signature of a non-sinusoidal CPR in the dependence of \( I_c(H) \).

Furthermore, since the maximum diode efficiency is almost independent of the asymmetry in \( J_c(x) \) but strongly depends on the CPR, measurements of \( \eta \) can serve as a tool to determine intrinsic properties of the Josephson junction, such as its transparency \( D \).

The substantially different behavior of $\eta$ within the finite-length junction approximation cannot be explained simply. This is due to a more complex interplay between two mechanisms: self-field and intrinsic. Nevertheless, Fig.~\ref{Fig:2}(c) shows that in the fairly long junction limit, the self-field mechanism surpasses the intrinsic one, whereas in the short junction regime, only the intrinsic mechanism is active. Moreover, the overall difference in the transition length within $1 < L/\lambda_J < 4$ of $\eta$ is approximately 25\%, which is close to the residual $\eta \simeq 30\%$ of the intrinsic mechanism between the two CPRs, highlighting this interplay.

To conclude, I examined the impact of the CPR on the properties of a single JJ in both short and finite-length regimes. In the short junction limit, it was shown that an asymmetric critical current density leads to a pronounced divergence in \( I_c(H) \), which is highly dependent on the CPR and manifests as nonreciprocity in the case of non-sinusoidal CPRs. The theoretical analysis of JDE in this regime, along with its comparison to the finite-length junction approximation, highlights that the effect originates from an intrinsic mechanism, independent of self-field contributions. Furthermore, it was demonstrated that nonreciprocity can arise even in JJs with minimal asymmetry, revealing a distinct fingerprint of the CPR on the critical current modulation. These findings suggest that JDE measurements could serve as a tool for probing intrinsic JJ parameters, such as the CPR.


\begin{acknowledgments}
The author thank Dr. Denis A. Bandurin, A. Kudriashov, Prof. Vladimir M. Krasnov and S. Grebenchuk  for insightful discussions and valuable feedback that contributed to this work.
\end{acknowledgments}


\begin{thebibliography}{99}
\bibitem{Barone}
Barone, A. \& Paterno, C. Physics and Applications of the Josephson Effect (J. Wiley \& Sons, New York, USA, 1982).

\bibitem{Likharev_1979}
Likharev, K.K.  Superconducting weak links.{\em Reviews of Modern Physics}, {\bf 51(1)},  101 (1979).

\bibitem{RSFQ}
Likharev, K.K. and Semenov, V.K. RSFQ logic/memory family: A new Josephson-junction technology for sub-terahertz-clock-frequency digital systems. IEEE transactions on applied superconductivity, {\bf 1(1)}, 3-28 (1991).

\bibitem{Bairamkulov_2025}
Bairamkulov, R. and De Micheli, G.,  Superconductive Electronics: A 25-Year Review . {\em IEEE Circuits and Systems Magazine}, {\bf 24(2)}, 16-33 (2024).
\bibitem{Larkin_1969}
Aslamazov, L. G., \& Larkin, A. I. Josephson effect in superconducting point contacts. { \em JETP Lett. }, {\bf 9}, 87 (1969).

\bibitem{Josepshon_1964}
Josephson, B.D. Coupled superconductors. {\em Reviews of Modern Physics}, {\bf 36(1)}, 216 (1964).

\bibitem{Kulik_Clean_1} 
Kulik, I.O.  Macroscopic quantization and the proximity effect in SNS junctions. {\em JETP Lett.}, {\bf30}, 944 (1969).

\bibitem{Kulik_Clean_2}
Kulik, I.O. and Omel'Yanchuk, A.N., . Properties of superconducting microbridges in the pure limit.{\em Soviet Journal of Low Temperature Physics}, {\bf 3(7)}, 459-461 (1977).

\bibitem{Kulik_Dirt}
Kulik, I.O. and Omelyanchouk, A.N.  Current flow in long superconducting junctions. {\em Sov. Phys. JETP} {\bf 41}, 1071 (1975).

\bibitem{Kulik_Phase}
Haberkorn, W., Knauer, H. and Richter, J. A theoretical study of the current‐phase relation in Josephson contacts. {\em Physica Status Solidi A}, {\bf 47(2)}, K161-K164 (1978).

\bibitem{Golubov_Review}
Golubov, A.A., Kupriyanov, M.Y. and Il’Ichev, E. The current-phase relation in Josephson junctions. {\em Reviews of Modern Physics}, {\bf 76(2)}, 411 (2004).

\bibitem{Zaikin}
Galaktionov, A.V. and Zaikin, A.D. Quantum interference and supercurrent in multiple-barrier proximity structures. {\em Physical Review B}, {\bf 65(18)}, 184507 (2002).

\bibitem{Ivanov_1981}
Ivanov, Z.G., Kupriyanov, M.Y., Likharev, K.K., Meriakri, S.V. and Snigirev, O.V. Boundary conditions for the Eilenberger and Usadel equations and properties of “dirty” SNS sandwiches. {\em Soviet Journal of Low Temperature Physics}, { \bf 7(5)}, 274-281 (1981). 

\bibitem{Kupr_1992}
Kupriyanov, M.Y. Effect of a finite transmission of the insulating layer on the properties of SIS tunnel junctions. {\em JETP Lett}, {\bf 56},414 (1992).

\bibitem{Buzdin_1982}
Buzdin, A.I. and Kupriyanov, M.Y. Josephson junction with a ferromagnetic layer.{\em JETP lett}, {\bf 53(6)}, 321 (1991).

\bibitem{Bezryadin_2008}
Dinsmore, R.C., Bae, M.H. and Bezryadin, A. Fractional order Shapiro steps in superconducting nanowires. {\em Applied physics letters},{\bf 93(19)}, 192505. (2008).

\bibitem{Troeman_2008}
Troeman, A.G.P., Van Der Ploeg, S.H.W., Il’Ichev, E., Meyer, H.G., Golubov, A.A., Kupriyanov, M.Y. and Hilgenkamp, H.,  Temperature dependence measurements of the supercurrent-phase relationship in niobium nanobridges. {\em Physical Review B}, {\bf 77(2)}, 024509 (2009).

\bibitem{Nanda_2017}
Nanda, G.; Aguilera-Servin, J. L.; Rakyta, P.; Kormányos, A.;
Kleiner, R.; Koelle, D.; Watanabe, K.; Taniguchi, T.; Vandersypen, L.
M.; Goswami, S. Current-phase relation of ballistic graphene Josephson
junctions. {\em Nano Lett.}  {\bf 17}, 3396-3401 (2017).

\bibitem{Stolyarov_2020}
Stolyarov, V.S., Yakovlev, D.S., Kozlov, S.N., Skryabina, O.V., Lvov, D.S., Gumarov, A.I., Emelyanova, O.V., Dzhumaev, P.S., Shchetinin, I.V., Hovhannisyan, R.A. and Egorov, S.V.,  Josephson current mediated by ballistic topological states in Bi2Te2. 3Se0. 7 single nanocrystals. {\em Communications Materials}, {\bf 1(1)}, 38 (2020).

\bibitem{Karter_2015}
Kurter, C., Finck, A.D., Hor, Y.S. and Van Harlingen, D.J.,  Evidence for an anomalous current–phase relation in topological insulator Josephson junctions. {\em Nature communications}, {\bf 6(1)}, 7130 (2015).

\bibitem{Spanton_2017}
Spanton, E.M., Deng, M., Vaitiekėnas, S., Krogstrup, P., Nygård, J., Marcus, C.M. and Moler, K.A. Current–phase relations of few-mode InAs nanowire Josephson junctions. {\em Nature Physics}, {\bf 13(12)}, 1177-1181 (2017).

\bibitem{Hart_2019}
Hart, S., Cui, Z., Ménard, G., Deng, M., Antipov, A.E., Lutchyn, R.M., Krogstrup, P., Marcus, C.M. and Moler, K.A. Current-phase relations of InAs nanowire Josephson junctions: From interacting to multimode regimes. {\em Physical Review B}, {\bf 100(6)}, 064523 (2019).

\bibitem{Frolov_2004}
Frolov, S.M., Van Harlingen, D.J., Oboznov, V.A., Bolginov, V.V. and Ryazanov, V.V. Measurement of the current-phase relation of superconductor/ferromagnet/superconductor $\pi$ Josephson junctions. {\em Physical Review B}, {\bf 70(14)}, 144505 (2004).

\bibitem{Sochnikov_2013}
Sochnikov, I., Bestwick, A.J., Williams, J.R., Lippman, T.M., Fisher, I.R., Goldhaber-Gordon, D., Kirtley, J.R. and Moler, K.A., Direct measurement of current-phase relations in superconductor/topological insulator/superconductor junctions. {\em Nano letters}, {\bf 13(7)}, 3086-3092 (2013).

\bibitem{Kayyalha_2020}
Kayyalha, M., Kazakov, A., Miotkowski, I., Khlebnikov, S., Rokhinson, L.P. and Chen, Y.P. Highly skewed current–phase relation in superconductor–topological insulator–superconductor Josephson junctions. {\em npj Quantum Materials}, {\bf 5(1)}, 7 (2020).

\bibitem{Rocca_2007}
Della Rocca, M.; Chauvin, M.; Huard, B.; Pothier, H.; Esteve, D.;
Urbina, C. Measurement of the current-phase relation of superconducting
atomic contacts. {\em Physical review letters} {\bf 99}, 127005 (2007).

\bibitem{Ginzburg_2018}
Ginzburg, L.V., Batov, I.E.E., Bol’ginov, V.V.E., Egorov, S.V.E., Chichkov, V.I., Shchegolev, A.E.E., Klenov, N.V., Soloviev, I.I., Bakurskiy, S.V. and Kupriyanov, M.Y.,  Determination of the current–phase relation in Josephson junctions by means of an asymmetric two-junction SQUID. {\em JETP Lett.}, {\em 107}, 48-54 (2018).

\bibitem{Enders_2023}
Endres, M., Kononov, A., Arachchige, H.S., Yan, J., Mandrus, D., Watanabe, K., Taniguchi, T. and Schonenberger, C.,  Current–phase relation of a WTe2 Josephson junction. {\em Nano Letters}, {\bf 23(10)}, 4654-4659 (2023).

\bibitem{Babich_2023}
Babich, I., Kudriashov, A., Baranov, D. and Stolyarov, V.S.,  Limitations of the current–phase relation measurements by an asymmetric dc-SQUID. {\em Nano Letters}, {\bf 23(14)}, 6713-6719 (2023).


\bibitem{Willsch_2024}
Willsch, D., Rieger, D., Winkel, P., Willsch, M., Dickel, C., Krause, J., Ando, Y., Lescanne, R., Leghtas, Z., Bronn, N.T. and Deb, P. Observation of Josephson harmonics in tunnel junctions. {\em Nature Physics}, {\bf 20(5)}, 815-821 (2024).

\bibitem{Goldobin_2007}
Goldobin, E., Koelle, D., Kleiner, R. and Buzdin, A. Josephson junctions with second harmonic in the current-phase relation: Properties of $\phi$ junctions. {\em Physical Review B}, {\bf 76(22)}, 224523 (2007).









\bibitem{Raz_2024}
Hovhannisyan, R.A., Golod, T. and Krasnov, V.M., Superresolution magnetic imaging by a Josephson junction via holographic reconstruction of $I_c (H)$ modulation. {\em Physical Review Applied}, {\bf 20(6)}, 064012 (2023).




\bibitem{Hall_2}
Hart, S., Ren, H., Wagner, T., Leubner, P., Mühlbauer, M., Brüne, C., Buhmann, H., Molenkamp, L.W. and Yacoby, A.,  Induced superconductivity in the quantum spin Hall edge. {\em Nature Physics}, {\bf 10(9)}, 638-643 (2014).

\bibitem{Hall_1}
Bocquillon, E., Deacon, R.S., Wiedenmann, J., Leubner, P., Klapwijk, T.M., Brüne, C., Ishibashi, K., Buhmann, H. and Molenkamp, L.W.,  Gapless Andreev bound states in the quantum spin Hall insulator HgTe. {\em Nature Nanotechnology}, {\bf 12(2)}, 137-143 (2017).


\bibitem{Chen_2024}
Chen, P., Wang, J., Wang, G., Ye, B., Zhou, L., Wang, L., Wang, J., Zhang, W., Chen, W., Mei, J. and He, H. Asymmetric edge supercurrents in MoTe 2 Josephson junctions. {\em Nanoscale Advances}, {\bf 6(2)}, 690-696 (2024).

\bibitem{Fominov_2022}
Fominov, Y.V. and Mikhailov, D.S.,  Asymmetric higher-harmonic SQUID as a Josephson diode. {\em Physical Review B}, {\bf 106(13)}, 134514 (2022).


\bibitem{Kudriashov_2025}
Kudriashov, A., Zhou, X., Hovhannisyan, R.A., Frolov, A., Elesin, L., Wang, Y., Zharkova, E.V., Taniguchi, T., Watanabe, K., Yashina, L.A. and Liu, Z. Non-Reciprocal Current-Phase Relation and Superconducting Diode Effect in Topological-Insulator-Based Josephson Junctions. arXiv preprint arXiv:2502.08527 (2025).


\bibitem{Wakatsuki1}
Wakatsuki, R., Saito, Y., Hoshino, S., Itahashi, Y.M., Ideue, T., Ezawa, M., Iwasa, Y. and Nagaosa, N. Nonreciprocal charge transport in noncentrosymmetric superconductors. {\em Science advances}, {\bf 3(4)}, e1602390 (2017). 

\bibitem{Wakatsuki2}
Wakatsuki, R. and Nagaosa, N. Nonreciprocal current in noncentrosymmetric Rashba superconductors. {\em Physical Review Letters}, {\bf 121(2)}, 026601 (2018).


\bibitem{Kim_2024}
Kim, J.K., Jeon, K.R., Sivakumar, P.K., Jeon, J., Koerner, C., Woltersdorf, G. and Parkin, S.S. Intrinsic supercurrent non-reciprocity coupled to the crystal structure of a van der Waals Josephson barrier. {\em Nature Communications}, {\bf 15(1)}, 1120 (2024).

\bibitem{Hoshino_2018}
Hoshino, S., Wakatsuki, R., Hamamoto, K. and Nagaosa, N. Nonreciprocal charge transport in two-dimensional noncentrosymmetric superconductors. {\em Physical Review B}, {\bf 98(5)}, 054510 (2018).

\bibitem{Prox}
Van Loo, N., Mazur, G.P., Dvir, T., Wang, G., Dekker, R.C., Wang, J.Y., Lemang, M., Sfiligoj, C., Bordin, A., van Driel, D. and Badawy, G. Electrostatic control of the proximity effect in the bulk of semiconductor-superconductor hybrids. {\em Nature Communications}, {\bf 14(1)}, 3325 (2023).

\bibitem{Reveiew_diode}
Nadeem, M., Fuhrer, M.S. and Wang, X.,  The superconducting diode effect. {\em Nature Reviews Physics}, {\bf 5(10)}, 558-577 (2023).




\bibitem{Krasnov_In}
Krasnov, V.M. Josephson junctions in a local inhomogeneous magnetic field. {\em Physical Review B}, {\bf 101(14)}, 144507  (2020).

\bibitem{Krasnov_Diode}
Krasnov, V.M., Oboznov, V.A. and Pedersen, N.F. Fluxon dynamics in long Josephson junctions in the presence of a temperature gradient or spatial nonuniformity. {\em Physical Review B}, {\bf 55(21)}, 14486 (1997).


\bibitem{Vilegas_2003}
Villegas, J. E. et al. A superconducting reversible rectifier that controls the motion of magnetic flux quanta. {\em Science} {\bf 302}, 1188 (2003).


\bibitem{Moschalkov_2006} de Souza Silva, C. C., Van de Vondel, J., Morelle, M. \& Moshchalkov, V. V. Controlled multiple reversals of a ratchet effect. {\em Nature} {\bf 440}, 651 (2006).


\bibitem{Lustikova_2018} Lustikova, J. et al. Vortex rectenna powered by environmental fluctuations.
{\em Nat. Commun.} {\bf 10}, 1038 (2018).


\end{thebibliography}
\end{document}